\documentclass[aps,a4paper,12pt]{revtex4}
\pdfoutput=1
\usepackage{float}
\usepackage{fancyhdr}
\usepackage{color}
\usepackage{graphicx}
\usepackage{epsfig} 
\usepackage{amssymb}
\usepackage{amsmath,amsfonts}
\usepackage{mathdots}
\usepackage{mathtools}
\usepackage{booktabs}
\usepackage[pagebackref=true]{hyperref}
\usepackage{tikz}
\usepackage{eurosym}
\usetikzlibrary{shapes}
\usepackage{pgfplotstable}
\usepackage[footnotesize,singlelinecheck=off,justification=raggedright]{caption}
\usepackage{multirow}
\usepackage{faktor}
\def\bra#1{\mathinner{\langle{#1}|}}
\def\ket#1{\mathinner{|{#1}\rangle}}

\def\scprod#1#2{\mathinner{\langle{\,#1\,}|{\,#2\,}\rangle}}

\DeclareMathAlphabet{\mathbbmsl}{U}{bbm}{m}{sl}
{\catcode`\|=\active\gdef\Braket#1{\left<\mathcode`\|"8000\let|\bravert {#1}\right>}}
\def\bravert{\egroup\,\vrule\,\bgroup}
\def\Tr{\mathop{\mbox{\normalfont Tr}}\nolimits}

\def\ii{{\rm i\hskip 0.5mm}}
\def\dd{{\rm d}}
\def\ee{{\rm e}}

\def\CP{{\mathcal P}}
\def\CH{{\mathcal H}}

\def\CL{{\mathcal L}}

\def\NR{{\mathbb R}}

\def\comment#1{}

\def\ker{{\rm ker}}
\def\ran{{\rm ran}}

\begin{document}

\title{Canonical quantization of the electromagnetic field\\ in arbitrary $\xi$-gauge.\footnote{Dedicated to the memory of my dearest professor and friend, Krzysztof Gaw\c{e}dzki.}}

\author{Fernando Falceto}
 \affiliation{Departamento de F\'{\i}sica Te\'orica, Universidad de Zaragoza,
50009 Zaragoza, Spain}
\affiliation{Instituto de Biocomputaci\'on y F\'{\i}sica de Sistemas
Complejos (BIFI)}   
\affiliation{Centro de Astropart\'\i culas y F\'\i sica de Altas Energ\'\i as (CAPA),
50009 Zaragoza, Spain}
    


\begin{abstract} 
  We carry out the canonical quantization of the electromagnetic field in arbitrary $\xi$-gauge and compute its propagator. In this way we fill a gap in the literature and clarify some existing confusion about Feynman $\ii\epsilon$ prescription for the propagator of the electromagnetic field. We  also discuss the BRST quantization and investigate the apparent singularities present in the theory when the gauge parameter $\xi$ takes the value -1.  We find that this is a mere artifact due to the choice of basic modes and show that in the appropriate basis the commutation relations and the BRST transformation are, in fact, independent of the gauge parameter. The latter only appears as the coefficient of a BRST exact term in the Hamiltonian, which constitutes an extremely simple proof of the independence of any physical process on the gauge parameter $\xi$. 
\end{abstract}

\maketitle

\section{Introduction.}

A  most important lesson I learnt from Krzysztof's research and teaching is that there is no good Theoretical Physics without honest Mathematics. This belief is visible already in his very first contributions \cite{Gaw1}. As I could experience myself when working with him, he struggled (and made me struggle too) to derive physical results in a convincing (or rigorous) mathematical way, see for instance \cite{Gaw2,Gaw3,Gaw4,Gaw5,Gaw6} as some paradigmatic examples of this in a great variety of areas in Mathematics and Physics. 

This intransigent attitude is certainly most valuable for the advance of Theoretical Physics, since, as we all know, there are occasions in which some results are commonly accepted and even appear in text books, but they have not been properly derived. They are sometimes guessed or obtained in a non rigorous way. 

It is somehow surprising that we can encounter one of these instances in the realm of elementary free Quantum Field Theory: the propagator of the $U(1)$ gauge field $A_\mu$ in a general $\xi$-gauge. A problem perfectly well defined in mathematical terms whose proper solution I was not able to find in the literature.

The Lagrangian density in the Gupta-Bleuler formalism for arbitrary gauge parameter $\xi$ reads
\begin{equation}\label{gblag}
  \CL_{em}=-\frac14 F_{\mu\nu}F^{\mu\nu} -\frac1{2\xi} (\partial_\mu A^\mu)^2,\quad F_{\mu\nu}=\partial_\mu A_\nu-\partial_\nu A_\mu
\end{equation}
where I use the positive time signature for the Minkowski metric, i. e.
$g={\rm diag}(1,-1,-1,-1)$.

Actually, in their original papers \cite{Gupt,Bleu}, the authors consider the particular case $\xi=1$, nowadays known as Feynman gauge. This case is the most thoroughly studied in the literature, see for instance \cite{StrWig} for a rigourous treatment. Here we will consider an arbitrary value for the gauge parameter $\xi$, instead.

After quantizing the theory defined by the Lagrangian in (\ref{gblag}), one may derive the Feynman propagator
\begin{equation}\label{propa}
  \bra 0 T A_\mu(x) A_\nu(y)\ket 0=\lim_{\epsilon\to 0^+}\int_{\NR^4}
        \Pi_{\mu\nu}^\epsilon(p) \ee^{-\ii\! p(x-y)} \frac{\dd^4p}{(2\pi)^4}.
\end{equation}
If you check the literature you will find that, starting at least from de classical review papers \cite{AbeLee,MarPag} and continuing with the majority of most popular text books \cite{Kaku,PesSch,Wein,Sred,Bank,Zeid,Schw}, the propagator in momentum space is written as
\begin{equation}\label{propap1}
  \Pi_{\mu\nu}^\epsilon(p)=\frac{-\ii}{p^2+\ii\epsilon}\left( g_{\mu\nu} - (1-\xi)\frac{p_\mu p_\nu}{p^2}\right).
\end{equation}
An expression that when inserted into (\ref{propap1}), and due to the poles at $p_0=\pm |\vec p|$
in the $\xi$ dependent term, produces an undefined or divergent result.
Hence, it cannot be considered as a valid expression for the propagator in momentum space.

Notably, in \cite{Wein} it is suggested to derive the expression for $\Pi_{\mu\nu}^\epsilon$ by inverting the quadratic form in the action with the $\ii \epsilon$ term included. Namely, solving for $\Pi_{\nu\rho}^\epsilon(p)$ in 
\begin{equation}\label{weinprop}
  \ii\left(g^{\mu\nu} (p^2+\ii\epsilon)
  - (1-\xi^{-1}){p^\mu p^\nu}\right) \Pi_{\nu\rho}^\epsilon(p)=\delta_\rho^\mu.
\end{equation}
From which one gets 
$$ \Pi_{\mu\nu}^\epsilon(p)=\frac{-\ii}{p^2+\ii\epsilon}\left( g_{\mu\nu} - (1-\xi)\frac{p_\mu p_\nu}{p^2+\ii\xi\epsilon}\right),$$
that, for $\xi\not=0$, cures the pole ambiguity leading to a well defined propagator in $x$-space. Notice, however, that for $\xi=0$ the problem persists and, moreover,  the location of the poles depends on $\xi$: if $\xi>0$ two poles are under the positive $p_{0}$ real line and two more over the negative one, then the standard (anticlock wise) Wick rotation is allowed; but if $\xi<0$ we have one pole on each side of the positive real line and the same in the negative one, hence the Wick rotation that leads to the euclidean formulation of the theory is not permitted any more.
It is surprising that a parameter that is supposed to have no physical meaning affects so deeply the consistency of the theory.

Actually, the proposal (\ref{weinprop}) has not a very solid justification. It is introduced just for solving the pole ambiguity and because it leads to the right result for the scalar and the gauge field in Feynman gauge ($\xi=1$).

In \cite{ItzZub} Claude Itzykson and Jean-Bernard Zuber carry out a sounder derivation of the propagator within the canonical quantization formalism. The authors follow an indirect procedure, though. They consider the massive Proca theory with the $\xi$-term added and after quantizing and computing the propagator they take the massless limit. Their result is
\begin{equation}\label{IZprop}
  \Pi_{\mu\nu}^\epsilon(p)=\frac{-\ii}{p^2+\ii\epsilon}
  \left( g_{\mu\nu} - (1-\xi)\frac{p_\mu p_\nu}{p^2+\ii\epsilon}\right),
\end{equation}
that, as we will see later, agrees with the right expression. This form of the propagator, without further proof, is also collected in \cite{Coll,Ster}

There are, however, two weak points in this derivation. First, it follows the slippery procedure of breaking the gauge symmetry by the addition of a mass term and then going to the limit in which the symmetry is restored.
In second place for $\xi<0$  it has tachyonic particles
(the authors already notice this and restrict themselves
to positive values of the gauge parameter).
In fact, the equations of motion for the theory are
$$(\partial_\mu\partial^\mu+m^2)A_\nu-(1-\xi^{-1})\partial_\nu\partial^\mu A_\mu=0$$
whose plane wave solutions have momenta $p^2=m^2$, with three polarisation vectors $q$ orthogonal to $p$ and one more solution polarized in the direction of $p$  with $p^2=\xi m^2$.
Then if $\xi<0$ the $p$-polarized mode is tachyonic and the causality issue appears again in this case. This seems to reinforce the idea that there may be some subtlety or causal violation when $\xi<0$. Which, as we will show, is not the case.

A third approach is followed by more metric-minded people \cite{CanRai,Ivas1,Ivas2,Ivas3,Viss1,Viss2}.
It consists in interpreting Feynman's $\ii\epsilon$ prescription as a complex deformation of the Minkowski metric, or more generally a pseudo Riemannian one.
In our case of interest and in order to account for the $\ii\epsilon$ term, we simply add an imaginary piece to the temporal part of the metric \cite{Viss1}
$$g^\epsilon_{\mu\nu}=g_{\mu\nu}+\ii\epsilon \delta_\mu^0\delta^0_\nu$$
so that we must replace $p^2$ in the propagator with
$$g^\epsilon_{\mu\nu}p^\mu p^\nu=p^2+\ii\epsilon p_0^2$$
that in the $\epsilon\to 0^+$ limit is equivalent to (\ref{IZprop}).
Of course, the previous deformation of the metric is linked to 
Wick rotation (see also \cite{LouSor} that illustrates the  use of complex metrics to induce topology change and more recently \cite{KonSeg} and \cite{Witt} in connection with Quantum Field Theory and Quantum Gravity respectively).
Anyhow, we should keep in mind, as emphasized by Matt Visser in \cite{Viss2} who I quote here, that ``Feynman's $\ii\epsilon$ prescription ... was originally developed as a pragmatic trick for encoding causality'' in the  propagator, and it is this prescription which ``justifies flat-space Wick rotation'', and not the other way around.

Hence in order to solve the puzzle, and according to the previous remark,
one must find the Feynman propagator for arbitrary $\xi$-gauge following
the standard pathway of Quantum Field Theory and then
try to implement the $\ii\epsilon$ ``pragmatic trick''.
I have seen the exercise proposed in different text books but I did not find its solution in the physics literature, the goal of this paper is to remedy this omission.
We will complete the paper with an account of the BRST quantization
of the theory.

\section{Canonical quantization in $\xi$-gauge}  

In order to quantize a free field theory one usually looks for the normal modes
(plane wave solutions) whose amplitudes satisfy, after applying the canonical
quantization prescriptions,  the commutation relations of creation and
annihilation operators.
From these we construct the Fock space of states for the free theory,
the Hamiltonian and all the relevant operators.

In our case the equations of motion are
\begin{equation}\label{eqmot}
  \partial_\mu\partial^\mu A_\nu+
  \frac{1-\xi}\xi\partial_\nu\partial^\mu A_\mu=0.
\end{equation}

At this point one usually takes $\xi=1$ (Feynman gauge) that, as we said, was the original proposal of Gupta and Bleuler\cite{Gupt,Bleu}. Of course, the reason for that choice is that in the Feynman gauge the equations have a basis of plane wave solutions of the form
\begin{equation}\label{planewave}
  \tilde u_\mu(x)={k}_\mu\ee^{-\ii p x},\quad p^2=0,
\end{equation}
where ${k}$ is an arbitrary polarization vector. It is customary
to consider for every $p$ the basis of four polarization modes:
temporal ${k}^{(0)}=(1,\vec 0)$, longitudinal ${k}^{(3)}=(0,\vec p/|\vec p|)$
and the two transverse ones ${k}^{(j)}=(0,\vec {k}^{\,(j)})$, $j=1,2$ with $\vec {k}^{\,(j)}\cdot\vec p=0$ and $\vec {k}^{\,(j)}\cdot \vec {k}^{\,(j')}=\delta_{jj'}$.

For $\xi\not=1$, however, if we plug the plane wave solution into (\ref{eqmot})
we obtain the necessary condition
\begin{equation}\label{eqmotmom}
  p^2{k}_\nu+\frac{1-\xi}\xi (p\cdot {k}) p_\nu=0.
\end{equation}
Which, multiplying it by $p^\nu$ and contracting indices, leads to
$$p^2 (p\cdot {k})=0,$$
but if $p^2=0$ and $\xi\not=1$ we derive from (\ref{eqmotmom}) $p\cdot {k}=0$
and vice-versa.

Hence, the polarization vector $k$ should satisfy the additional condition of being orthogonal to $p$. While transverse modes above are still solutions of the equations of motion, the temporal and
longitudinal modes are not, only their sum survives. It seems that
there is a mode missing.

To find the hidden mode we must modify the ansatz. The situation is
reminiscent of the resonance in an harmonic oscillator
where solutions of the form $t\cos(\omega t)$ appear.

Actually, one can straightforwardly check that
\begin{equation}\label{newsol}
  u_\mu(x)=\left({k}_\mu+\ii\frac{1-\xi}{1+\xi} ({k}\cdot x) p_\mu\right)\ee^{-\ii p x},\quad
p=(|\vec p|,\vec p),
\end{equation}
is a solution of (\ref{eqmot}) for any ${k}$.
The idea of adding an $x$-linear term
to the polarization vector is already mentioned
in \cite{ItzZub}, although formulated in a somehow vague or cryptic way.
Referring to the solution of (\ref{eqmot}), the authors write:
``Necessarily, [it] involves $\delta(p^2)$ and $\delta'(p^2)$''.
Ours is a precise implementation of this idea. 

Now, we can use these solutions to complete the basis of modes. In doing so
one must keep in mind that the goal is to perform the canonical quantization
at constant time, so it would be advisable to look for a basis which is
orthogonal (or rather $\delta$-orthogonal) when integrated in space. This requirement conflicts with the
$x$ dependence of the polarization vector in (\ref{newsol}), except if we
choose its temporal component $x^0$ or, in other words, except if we take
${k}={k}^{(0)}=(1,\vec 0)$ in the new solutions.

To be precise we will consider the following basis of solutions
\begin{eqnarray*}
  u_{\vec p}^{(0)}(x)&=&({k}^{(0)}+\ii\frac{1-\xi}{1+\xi} x^0 p)\ee^{-\ii p x}\cr
  u_{\vec p}^{(3)}(x)&=&({k}^{(3)}-\ii\frac{1-\xi}{1+\xi} x^0 p)\ee^{-\ii p x}=|\vec p|^{-1}p\,\ee^{-\ii p x}-u_{\vec p}^{(0)}(x)\cr
  u_{\vec p}^{(j)}(x)&=&{k}^{(j)}\ee^{-\ii p x},\quad j=1,2
\end{eqnarray*}
with ${k}^{(\lambda)}, \lambda=0,1,2,3$, the orthogonal vectors that we introduced
before and $p=(|\vec p|,\vec p)$.

Defining new (time dependent) {\it polarization} vectors, $\epsilon^{(\lambda)}(\vec p,x^0)$, we can write more compactly:
  $$u_{\vec p}^{(\lambda)}(x)=\epsilon^{(\lambda)}(\vec p,x^0)\ee^{-\ii p x}\quad
  \mbox{with}\quad \epsilon^{(\lambda)}(\vec p,x^0)={k}^{(\lambda)}+\ii(g^{\lambda0}+g^{\lambda3})\frac{1-\xi}{1+\xi} x^0 p.$$
  Notice that this basis reduces to the standard one when $\xi=1$, making therefore possible to compare our results with those in the Feynman gauge.

  The new polarization vectors satisfy (complex) orthonormality and completeness relations, namely
  $$\epsilon_\mu^{(\lambda)}(\vec p,x^0)^*\epsilon_\nu^{(\lambda')}(\vec p,x^0)g^{\mu\nu}=
  g^{\lambda\lambda'},\qquad
    \epsilon_\mu^{(\lambda)}(\vec p,x^0)^*\epsilon_\nu^{(\lambda')}(\vec p,x^0) g_{\lambda\lambda'}=
    g_{\mu\nu},
    $$
    which implies
    $$\int_{\NR^3}u_{\vec p,\mu}^{(\lambda)}(x)^*
    u_{\vec q,\nu}^{(\lambda')}(x)g^{\mu\nu}\,{\dd^3 x}=
    (2\pi)^3g^{\lambda\lambda'}\delta^3(\vec p-\vec q),
    $$
    $$
    \int_{\NR^3}u_{\vec p,\mu}^{(\lambda)}(x^0,\vec x)^*
    u_{\vec p,\nu}^{(\lambda')}(x^0,\vec y)g_{\lambda\lambda'}\,{\dd^3 p}=
    (2\pi)^3g_{\mu\nu}\delta^3(\vec x-\vec y).
    $$
That are the conditions for having a complete orthonormal basis of modes at fixed time, as we sought.

 The next step in the standard quantization procedure for free theories is to write the field as a superposition of these modes, 
 $$A_\mu(x)=\int\sum_{\lambda=0}^3(a_{\vec p}^{(\lambda)}\epsilon_\mu^{(\lambda)}(\vec p,x^0)\ee^{-\ii px}
 +
 {a_{\vec p}^{(\lambda)}}^\dagger{\epsilon_\mu^{(\lambda)}}(\vec p,x^0)^*\ee^{\ii px})\frac{\dd^3 p}{(2\pi)^3\sqrt{2|\vec p|}},
 $$
 where  ${a_{\vec p}^{(\lambda)}}^\dagger$ is the adjoint operator of
   $a_{\vec p}^{(\lambda)}$ which guarantees the hermiticity of the field $A_\mu(x)$.

 Its inverse relation reads
 $$\sum_{\lambda=0}^3a_{\vec p}^{(\lambda)}\epsilon_\mu^{(\lambda)}(\vec p)
= \int(\ii \dot A_\mu +|\vec p| A_\mu+\ii\frac{1-\xi}{2\xi}
 \frac{p_\mu}{|\vec p|}\partial_\nu A^\nu)\ee^{\ii px}\frac{\dd^3x}{\sqrt{2|\vec p|}}.$$

 Using the canonical commutation relations between the fields $A_\mu$ and
 their momenta
 $$\pi^0=-\frac1\xi(\dot A^0+\vec\nabla\vec A),\quad \vec \pi=\dot{\vec A}+\vec\nabla A^0$$
 and the previous expressions relating the gauge field and  the creation annihilation operators, one can derive the commutation relations of the latter.
 The way we computed them involves elementary analysis but it is a little
 tedious, though.
 We give here only the final result that, by the way, is so simple that it
 should deserve a more direct way to derive it. Our final
 result is that the only non vanishing commutation relations are
 \begin{eqnarray}
   &&[{a_{\vec p}^{(0)}},{a_{\vec q}^{(0)}}^\dagger]=-(2\pi)^3
\frac{1+\xi}2
      \delta^3(\vec p-\vec q)\cr
      &&[{a_{\vec p}^{(3)}},{a_{\vec q}^{(3)}}^\dagger]=\ \ (2\pi)^3
\frac{1+\xi}2 \delta^3(\vec p-\vec q)\\[1mm]
   &&[{a_{\vec p}^{(j)}},{a_{\vec q}^{(j')}}^\dagger]=\ 
   (2\pi)^3
   \delta^3(\vec p-\vec q)\delta_{jj'},\ j,j'=1,2\nonumber.
 \end{eqnarray}

 Once we have determined the commutation relations of the creation
 annihilation  operators we can construct the space of states in the usual way: the Fock space obtained from the vacuum $\ket{\Omega}$, defined as the
   only state killed by all annihilation operators, by acting on it with the creation operators.
   For instance, one particle states are $\ket{\vec p,\lambda}=\sqrt{2|\vec p|}{a_{\vec p}^{(\lambda)}}^\dagger\ket{\Omega}$, where we adopt the usual Lorentz invariant normalization. In fact, from the commutation relations and the properties of the vacuum one has
   $$\scprod{\vec p, j}{\vec q,j'}= 2|\vec p| (2\pi)^3\delta^3(\vec p-\vec q)\delta_{jj'},\quad j,j'=1,2$$
for the transverse modes, while for the temporal and longitudinal ones ($\lambda,\lambda'=0\ \mbox{or}\ 3$) we get
   $$\scprod{\vec p, \lambda}{\vec q,\lambda'}= (1+\xi)|\vec p| (2\pi)^3\delta^3(\vec p-\vec q){\mathbf J}_{\lambda\lambda'},
   \ \mbox{}\ \ 
   \mathbf J=\begin{pmatrix}-1&0\cr0&1\end{pmatrix},$$
  i. e. the scalar product is of the indefinite type. 

In the Feynman gauge, $\xi=1$, and more generally for $\xi>-1$,
  we have the standard scenario in which the temporal creation operators produce negative square norm states. However for $\xi<-1$ the situation is inverted: temporal states have a positive square norm while longitudinal states (generated by $a_{\vec p}^{(3)}$) have a negative one. Also when $\xi=-1$ we have that both, temporal and longitudinal states, have zero norm and the scalar product is degenerate. One may wonder if there is some singularity in the theory when $\xi=-1$ or it is an artifact of our procedure.

 This is somehow reminiscent of what happens at the black hole horizon where the radial and temporal directions exchange there signatures and the metric is singular. In that case the singularity is due to the choice of coordinates and there is nothing physically meaningful behind. As we shall see later we have here exactly the same situation. But, for now, let us proceed to compute other relevant operators of the theory like the Hamiltonian and the momentum.

 We consider in first place the normal ordered Hamiltonian $H$, which is
 derived in the standard way from the Gupta-Bleuler Lagrangian (\ref{gblag})
 and can be computed
integrating in space the following density
 \begin{eqnarray}
   \CH=\frac12 
 :\dot {\vec A}^2-(\vec\nabla A^0)^2-(\vec\nabla\times\vec A)^2:
   -\frac1{2\xi}: (\dot A^0)^2+(\vec\nabla\vec A)^2:
 \end{eqnarray}
 After performing the integration we can write $H$ in terms of the creation annihilation operators to give
 \begin{eqnarray}
H&=&\int|\vec p|({a^{(1)}_{\vec p}}^\dagger a^{(1)}_{\vec p}+
 {a^{(2)}_{\vec p}}^\dagger a^{(2)}_{\vec p})\frac{\dd^3 p}{(2\pi)^3}\\
&+&\frac2{(1+\xi)^2}\int|\vec p|
 \left(({a^{(3)}_{\vec p}}^\dagger-{a^{(0)}_{\vec p}}^\dagger)(a^{(3)}_{\vec p}+
   \xi a^{(0)}_{\vec p})+
   ({a^{(3)}_{\vec p}}^\dagger+\xi {a^{(0)}_{\vec p}}^\dagger)
   (a^{(3)}_{\vec p}-a^{(0)}_{\vec p})\right)\frac{\dd^3 p}{(2\pi)^3}.\nonumber
 \end{eqnarray}
 Notice that in the Feynman gauge $\xi=1$ the expression simplifies notably and
 we obtain the standard result
 $$H_{\xi=1}=-\int|\vec p|{a^{(\lambda)}_{\vec p}}^\dagger a^{(\lambda')}_{\vec p}
 g_{\lambda\lambda'}\frac{\dd^3 p}{(2\pi)^3}.$$

 Going back to the general case, if we compute the action of the Hamiltonian
 on the transverse modes we obtain the expected result
 $H\ket{\vec p,j}=|\vec p|\ket{\vec p,j},\ j=1,2$;
 but when we do the same for the temporal and longitudinal modes we have some
 surprise. In fact,
 $$
 \begin{pmatrix}
   H\ket{\vec p,0}
   \cr
   H\ket{\vec p,3}
 \end{pmatrix}=|\vec p|\;
 \mathbf M
 \begin{pmatrix}
   \ket{\vec p,0}
   \cr
   \ket{\vec p,3}
 \end{pmatrix},\quad
\mathbf M=\frac1{1+\xi}\begin{pmatrix}
  2\xi&1-\xi\cr\xi-1&2
\end{pmatrix}.
$$
But notice that $\det\mathbf M=1$ and $\Tr\mathbf M=2$, which implies that
$\mathbf M$ is non-diagonalizable (except if it is the identity which occurs for $\xi=1$, the Feynman gauge). This means that there is not a basis of states with energy well defined, indeed one can check that the only eigenstate of $H$ in the $0,3$ polarization space is
$\ket{\vec p,0}+\ket{\vec p,3}$ with eigenvalue $|\vec p|$.
One may wonder how this is possible given that $H$ is a Hermitian operator. The apparent contradiction is solved if we recall that the scalar product in the space of longitudinal and temporal modes is indefinite or, in other words, Hermiticity of $H$ implies
$$\mathbf M^\dagger \mathbf J=\mathbf J\mathbf M$$
which does not guarantee the diagonalizability of $\mathbf M$.

Moving on to the momentum operator, associated to the invariance under spatial translations, it can be obtained from the density
 $$
 \vec\CP = : \dot A_\mu\vec\nabla A^\mu+
 \frac{1-\xi}\xi(\dot{A^0}+\vec\nabla\vec A)
 \vec\nabla A^0:
 $$ 
 and, once performed the spatial integration we obtain the momentum operator
 $$
 \vec P=- \int\vec p\, {a^{(\lambda)}_{\vec p}}^\dagger
 a^{(\lambda')}_{\vec p} g_{\lambda\lambda'}
 \frac{\dd^3 p}{(2\pi)^3}+
 \frac{1-\xi}{1+\xi}
 \int\vec p ({a^{(3)}_{\vec p}}^\dagger a^{(3)}_{\vec p}
 -{a^{(0)}_{\vec p}}^\dagger a^{(0)}_{\vec p})
 \frac{\dd^3 p}{(2\pi)^3}.
 $$
 We have no surprises in this occasion and the momentum of one particle states is well defined and as expected
 $$\vec P\ket{\vec p,\lambda}
 =\vec p \ket{\vec p,\lambda}.
 $$
 
 After this brief account of the quantization of the theory
 we proceed to fulfill our initial goal.

 \section{Feynman propagator}

 We are now in position to compute the Feynman propagator,
 i. e. the expectation value in vacuum of the time ordered product of gauge fields, 
   $$ D_{\mu\nu}(x,y)=\langle\Omega|TA_\mu(x)A_\nu(y)|\Omega\rangle.$$
    As this result is the main goal (and probably the main contribution)
    of the paper we will present it with certain detail.

    Expanding the gauge fields in terms of the creation annihilation
    operators we have
    \begin{eqnarray}\label{propa}
      &&\hskip -3mm
      D_{\mu\nu}(x,y)\\[1mm]
      &&\hskip -3mm
      =\theta(x^0\!\!-\!y^0)\!\int\!\!\!\int\!
      \sum_{\lambda,\lambda'}
      \epsilon_\mu^{(\lambda)}(\vec p, x^0)
      \epsilon_\nu^{(\lambda')}(\vec q, y^0)^*\,
      \langle\Omega|
      a_{\vec p}^{(\lambda)} {a_{\vec q}^{(\lambda')}}^\dagger
      |\Omega\rangle
      \ee^{-\ii(px-qy)}
     {\frac{\dd^3 q}
           {(2\pi)^3\sqrt{2|\vec q|}}
     \frac{\dd^3 p}
           {(2\pi)^3\sqrt{2|\vec p|}}}\nonumber\\
     &&\hskip -3mm
     +\theta(y^0\!\!-\!x^0)\!\int\!\!\!\int\!
      \sum_{\lambda,\lambda'}
      \epsilon_\nu^{(\lambda')}(\vec q, y^0)
      \epsilon_\mu^{(\lambda)}(\vec p, x^0)^*\,
      \langle\Omega|
      a_{\vec q}^{(\lambda')} {a_{\vec p}^{(\lambda)}}^\dagger
      |\Omega\rangle
      \ee^{-\ii(qy-px)}
     {\frac{\dd^3 q}
           {(2\pi)^3\sqrt{2|\vec q|}}
     \frac{\dd^3 p}
           {(2\pi)^3\sqrt{2|\vec p|}}}\nonumber
    \end{eqnarray}
where we have omitted terms containing expectation values like
$\langle\Omega|{a_{\vec p}^{(\lambda)}} {a_{\vec q}^{(\lambda')}}|\Omega\rangle$,
$\langle\Omega|{a_{\vec p}^{(\lambda)}}^\dagger {a_{\vec q}^{(\lambda')}}|\Omega\rangle$
and
$\langle\Omega|{a_{\vec p}^{(\lambda)}}^\dagger
{a_{\vec q}^{(\lambda')}}^\dagger|\Omega\rangle$
that always vanish.

From the commutation relations we obtain
$$ \langle\Omega|
      a_{\vec p}^{(\lambda)} {a_{\vec q}^{(\lambda')}}^\dagger
      |\Omega\rangle=(2\pi)^3\left(-g_{\lambda\lambda'}+\frac{1-\xi}2(g_{\lambda0}g_{\lambda'0}
      -g_{\lambda3}g_{\lambda'3})\right)\delta^3(\vec p-\vec q)$$
and using the $\delta$-function we can particularize to the case $\vec q=\vec p$, which we do in the following.

      The completeness relations, for two different times, acquire an extra term
      $$\epsilon_\mu^{(\lambda)}(\vec p, x^0)
      \epsilon_\nu^{(\lambda')}(\vec p, y^0)^*g_{\lambda\lambda'}
=g_{\mu\nu}+\ii\frac{1-\xi}{1+\xi}(x^0-y^0)\frac{p_\mu p_\nu}{|\vec p|},$$
where the identity  $k^{(0)}+k^{(3)}=p/|\vec p|$ was used.
By the same token, removing the transverse modes from the sum, we get
$$\epsilon_\mu^{(\lambda)}(\vec p, x^0)
      \epsilon_\nu^{(\lambda')}(\vec p, y^0)^*(g_{\lambda0}g_{\lambda'0}
      -g_{\lambda3}g_{\lambda'3})= {k}_\mu^{(0)}{k}_\nu^{(0)}-{k}_\mu^{(3)}{k}_\nu^{(3)}
      +\ii\frac{1-\xi}{1+\xi}(x^0-y^0)\frac{p_\mu p_\nu}{|\vec p|}.
      $$

      Putting everything together we obtain
      \begin{equation}
       \sum_{\lambda,\lambda'}
      \epsilon_\mu^{(\lambda)}(\vec p, x^0)
      \epsilon_\nu^{(\lambda')}(\vec q, y^0)^*\,
      \langle\Omega|
      a_{\vec p}^{(\lambda)} {a_{\vec q}^{(\lambda')}}^\dagger
      |\Omega\rangle=(2\pi)^3 R_{\mu\nu}(\vec p,x^0-y^0)\delta^3(\vec p-\vec q)
            \end{equation}
with
\begin{equation*}
  R_{\mu\nu}(\vec p,t)=
      -g_{\mu\nu}+\frac{1-\xi}2\left(
        k_\mu^{(0)}k_\nu^{(0)}-k_\mu^{(3)}k_\nu^{(3)}-\ii t\,\frac{p_\mu p_\nu}{|\vec p|}\right)
\end{equation*}
and remember that we have defined $(p^\mu)=(|\vec p|,\vec p)$.

Therefore, the propagator (\ref{propa}) reads
\begin{eqnarray*}
  &&D_{\mu\nu}(x,y)=\int
        \Big(\ \theta(x^0 - y^0)R_{\mu\nu}(\vec p,x^0-y^0)\ee^{-\ii p(x-y)}\\
&&\hskip 2.8cm +\theta(y^0 - x^0)R_{\mu\nu}(\vec p,y^0-x^0)\ee^{-\ii p(y-x)}\Big)
       \frac{\dd^3 p}
            {(2\pi)^3 2|\vec p|}\nonumber
\end{eqnarray*}

The contribution of  the first term of $R_{\mu\nu}$, proportional to $g_{\mu\nu}$, can be obtained using the standard $\ii\epsilon$ Feynman trick, in fact
\begin{equation*}
  \lim_{\epsilon\to 0^+}\int_{-\infty}^\infty\frac{-i g_{\mu\nu}}{p^2+\ii\epsilon}\ee^{-\ii {{p^{}_{}}_{}}_{\!0}\,(x^0-y^0)}\frac{\dd p_0}{2\pi}
  =\frac{-g_{\mu\nu}}{2|\vec p|}
  \left(
  \theta(x^0-y^0)\ee^{-\ii |\vec p|(x^0-y^0)}
  +\theta(y^0-y^0)\ee^{-\ii |\vec p|(y^0-x^0)}
  \right),
\end{equation*}
where in the left hand side $p^2=p_0^2-|\vec p|^2$.

The other term, dependent on $\xi$, is of a
different nature. Actually, the presence of a term linear in the difference
of times suggests a derivative of the exponential which in turns implies
the existance of a double pole. If for definiteness we consider the case
$x^0>y^0$, then completing appropriately the integration  contour in the
lower complex half-plane and applying the method of residues one gets
$$
\lim_{\epsilon\to 0^+}\int_{-\infty}^\infty\frac{p_\mu p_\nu}{(p^2+\ii\epsilon)^2}\ee^{-\ii {{p^{}_{}}_{}}_{\!0}\,(x^0-y^0)}\frac{\dd p_0}{2\pi}
=
-\ii
\frac{\partial}{\partial p_0}\left[\frac{p_\mu p_\nu}{(p_0+|\vec p|)^2}\ee^{-\ii {{p^{}_{}}_{}}_{\!0}\,(x^0-y^0)}\right]_{p_0=|\vec p|}
$$
where, in the previous expressions I take
$(p_\mu)=(p_0,-\vec p)$.

Now observe that computing the derivative above one has
$$
\frac\partial{\partial p_0}\left[\frac{p_\mu p_\nu}{(p_0+|\vec p|)^2}\right]_{p_0=|\vec p|}
  =
  \begin{cases}
    \displaystyle
    \frac1{4|\vec p|}&\mu,\nu=0\cr
    0&\mu\not=0, \nu=0\cr
  \displaystyle
-\frac{p_\mu p_\nu}{4|\vec p|^3}&\mu,\nu\not=0
  \end{cases}\quad=\frac{k^{(0)}_\mu k^{(0)}_\nu- k^{(3)}_\mu k^{(3)}_\nu}{4|\vec p|},
  $$
and 
  $$
  \left[
    \frac{p_\mu p_\nu}{(p_0+|\vec p|)^2}
    \frac{\partial}{\partial p_0}
    \ee^{-\ii {{p^{}_{}}_{}}_{\!0}\,(x^0-y^0)}
    \right]_{p_0=|\vec p|} = -\ii(x^0-y^0)\frac{p_\mu p_\nu}{4|\vec p|^2}\ee^{-\ii |\vec p|(x^0-y^0)}
  $$
  where in the right hand side of the last expression we recover the on shell convention
  $(p_\mu)=(|\vec p|,-\vec p)$.
  
  From the previous identities, combined with the appropriate coefficients,
  one finally obtains the desired result
  $$D_{\mu\nu}(x,y)=
  \lim_{\epsilon\to 0^+}\int\frac{-\ii}{p^2+\ii\epsilon}
  \left( g_{\mu\nu} - (1-\xi)\frac{p_\mu p_\nu}{p^2+\ii\epsilon}\right)\ee^{-\ii p\,(x-y)}
  \frac{\dd^4 p}
  {(2\pi)^4},$$
which agrees with the propagator in \cite{ItzZub}, but it has been derived in this occasion following a quite orthodox path.

\section{BRST symmetry}

To give a more complete account of the quantization of the electromagnetic field for arbitrary gauge fixing parameter, $\xi$, and to study its unitarity,
it is convenient to incorporate the Faddeev-Popov ghosts and the
Bechi-Rouet-Stora-Tuytin (BRST) symmetry. This will be the content of the present section.

First we introduce the ghost and the antighost, a pair of real,  Grassmann, scalar fields $c$ and $\overline c$ that should be quantized with anticommutators
(violating therefore the spin-statistics correspondence, hence its name).
Its Lagrangian is
$$\CL_{gh} =\ii(\partial_\mu\overline c)(\partial^\mu c),$$
with equations of motion
$$\partial_\mu\partial^\mu c=0,\quad\partial_\mu\partial^\mu \overline c=0$$
and canonically conjugated momenta
$$\pi_c=\ii\dot{\overline c},\qquad \pi_{\overline c}=-\ii\dot c.$$
Notice the minus sign in the definition $\pi_{\overline c}$  which is due to the Grassmannian character of the classical fields.

The solutions of the equations of motion are plane waves $\exp(-\ii px),\ (p^\mu)=(|\vec p|,\vec p)$, and using them we expand the fields in modes in the
following convenient way
$$
c(x)= \int
(
\alpha_{\vec p} \ee^{-\ii px}
+
{\alpha_{\vec p}}^\dagger \ee^{\ii px}
)
\frac{\dd^3 p}{(2\pi)^3\sqrt{2|\vec p|}}
$$
$$
\overline c(x)= \int
(
\ii
\overline
\alpha_{\vec p} \ee^{-\ii px}
-
\ii
{\overline \alpha_{\vec p}}^\dagger \ee^{\ii px}
)
\frac{\dd^3 p}{(2\pi)^3\sqrt{2|\vec p|}}.
$$
Where the expansion is arranged to guarantee the Hermiticity of the quantum fields.

  From the canonical anticommutation relations between the fields and their momenta we deduce
  that the only non vanishing anticommutators of the new creation annihilation operators are
  $$
  \{\alpha_{\vec p},\overline\alpha_{\vec q}^\dagger\}=
\{\overline\alpha_{\vec p},\alpha_{\vec q}^\dagger\}=
(2\pi)^3\delta^3(\vec p -\vec q)
$$
while the latter ones commute with those of the gauge field
$a_{\vec p}$  and $a_{\vec p}^\dagger$.

Of course, the full Lagrangian
$$\CL= \CL_{em}+\CL_{gh}$$
possesses the rigid BRST symmetry given by
\begin{eqnarray}
&&\delta_{\eta} A_\mu(x)=-\ii \eta\,\partial_\mu c(x)\cr
&&\delta_{\eta} c(x)=0\cr
&&\delta_{\eta} \overline c(x)=\frac1\xi\eta\, \partial^\mu A_\mu(x)
\end{eqnarray}
where $\eta$ is a real Grassmann variable that parameterizes the transformation.
The meaning of the BRST symmetry is somehow more transparent if we write it in terms of the creation and annihilation operators
\begin{align}\label{BRST}
  &\delta_{\eta} a_{\vec p}^{(0)}
  =-\eta\,|\vec p| \alpha_{\vec p},
  \quad
  &&\delta_{\eta} {a_{\vec p}^{(0)}}^\dagger=\eta\,|\vec p| {\alpha_{\vec p}}^\dagger,\cr
  &\delta_{\eta} a_{\vec p}^{(3)}=-\eta\,|\vec p| \alpha_{\vec p},
  \quad
  &&\delta_{\eta} {a_{\vec p}^{(3)}}^\dagger=\eta\,|\vec p| {\alpha_{\vec p}}^\dagger,\\[1mm]
  &\delta_{\eta} \overline\alpha_{\vec p}=\eta\,\frac{2|\vec p|}{\xi+1}
  (a_{\vec p}^{(3)}-a_{\vec p}^{(0)}),
  &&\delta_{\eta} {\overline\alpha_{\vec p}}^\dagger=\eta\,\frac{2|\vec p|}{\xi+1}
  ({a_{\vec p}^{(3)}}^\dagger-{a_{\vec p}^{(0)}}^\dagger),\nonumber
\end{align}
and all the rest  ($a_{\vec p}^{(j)},\ j=1,2,\ \alpha_{\vec p}$\, and there adjoints)
do not transform. With the previous expressions it is immediate to see the 
nilpotency of the BRST symmetry, i. e. $\delta_\eta\delta_{\eta'}=0$.

The next step is to find the operator, $Q_{\rm BRST}$, that, acting on the Fock space of the ghosts and gauge fields, implements the symmetry. Equivalently, it  should induce the BRST transformations of (\ref{BRST}) with
$$\delta_\eta X=\eta\ [\hskip -2mm[\ Q_{\rm BRST},X\ ]\hskip -2mm]\, ,$$
where $\,[\hskip -2mm[\ \ ,\ \ ]\hskip -2mm]\,$ stands  for the commutator or the anticommutator
depending on whether $X$ has  even or odd ghost number.

One can check that the operator we seek is 
$$Q_{\rm BRST}=\frac2{\xi+1}\int |\vec p|\left(
(a_{\vec p}^{(3)} -a_{\vec p}^{(0)}) {\alpha_{\vec p}}^\dagger
+
\alpha_{\vec p}({a_{\vec p}^{(3)}}^\dagger -{a_{\vec p}^{(0)}}^\dagger) 
\right)
\frac{\dd^3p}{(2\pi)^3};
$$
that is self adjoint, nilpotent, commutes with the Hamiltonian
and annihilates the vacuum $\ket\Omega$.
The previous properties guarantee that we have a well defined dynamics in the
quotient space
$$\CH_{\rm phys}=\faktor{\ker(Q_{\rm BRST})}{\ran(Q_{\rm BRST})},$$
which correctly inherits a scalar product from the Fock space.
Moreover, one can check that the scalar product is positive definite
which provides the arena to build a sensible physical theory.
This is the reason why we denote by $\CH_{\rm phys}$ the final Hilbert space.

There is still a problem with our construction: as we already mentioned
before, some singularities happen when $\xi=-1$. For instance the
BRST operator diverges. The question again is whether this is
something essential in the Gupta-Bleuler quantization or it is a mere artifact
of our procedure. It turns out that the second possibility holds and it is
possible to carry out the quantization without any  singular value for $\xi$,
except $\xi=\infty$ which is the {\it essential} singularity (we refer to \cite{StrWig} for an interesting discussion of the origin of this singularity).
In the next section we will show how to proceed.

\section{Light-like polarized modes}

If we trace back the origin of the singularity at $\xi=-1$
it is immediately evident that there is already a problem in the
definition of the basic solutions $u_{\vec p}^{(0)}$ and $u_{\vec p}^{(3)}$:
both diverge when $\xi\to -1$.

It is also clear that we may cure the problem by changing the basis so
that it is valid for any finite $\xi$. One particularly useful choice is
to consider the new modes
\begin{eqnarray}
  &&v_{\vec p,\mu}=u_\mu^{(3)}+u_{\mu}^{(0)} =
  |\vec p|^{-1} p_\mu\ee^{-\ii px}\\[2mm]
  &&\overline v_{\vec p,\mu}=\frac{1+\xi}4(u_\mu^{(3)}-u_{\mu}^{(0)})=\left(\frac{1+\xi}4|\vec p|^{-1} \overline p_\mu
  -\ii \frac{1-\xi}2 x^0 p_\mu\right)\ee^{-\ii px},
\end{eqnarray}
where we define $\overline p$ as the negative energy partner of $p$, i. e.
$(\overline p^\mu)=(-|\vec p|,\vec p)$.
These solutions are well defined and linearly independent for any finite $\xi$.
Note, besides, that they are  light-like polarized, i. e. 
$$v_\mu^* v^\mu=\overline v_\mu^* \overline v^\mu=0.$$

We expand the field in terms of the new modes (together with the original transversal ones)
\begin{eqnarray}
  A(x)&=&\int\sum_{j=1,2}(a_{\vec p}^{(j)} u_{\vec p}^{(j)}(x)+
  {a_{\vec p}^{(j)}}^\dagger u_{\vec p}^{(j)}(x)^*)
\frac{\dd^3 p}{(2\pi)^3\sqrt{2|\vec p|}}\cr
 &+&
\int(
b_{\vec p}\; v_{\vec p}(x)+
\overline b_{\vec p}\; \overline v_{\vec p}(x)
+{b_{\vec p}}^\dagger v_{\vec p}(x)^*+
{\overline b_{\vec p}}^\dagger \overline v_{\vec p}(x)^*
)
\frac{\dd^3 p}{(2\pi)^3\sqrt{2|\vec p|}},
\end{eqnarray}
where we introduce the annihilation creation operators
$b_{\vec p},\, \overline b_{\vec p},\, {b_{\vec p}}^\dagger,\,{\overline b_{\vec p}}^\dagger$
associated to the light-like modes.
They are related to the temporal and longitudinal ones by
\begin{eqnarray}
  b_{\vec p}&=&\frac12(a^{(3)}_{\vec p}+a^{(0)}_{\vec p})\cr
  \overline b_{\vec p}&=&\frac2{1+\xi}(a^{(3)}_{\vec p}-a^{(0)}_{\vec p}),
\end{eqnarray}
and their non vanishing commutation relations are
$$[b_{\vec p}, {\overline b_{\vec q}}^\dagger]=[\overline b_{\vec p}, {b_{\vec q}}^\dagger]
= (2\pi)^3\delta^3(\vec p- \vec q)
$$
as can be easily deduced from the commutation relation of $a^{(0)},
a^{(3)}$ and their adjoints.

If we rewrite the BRST transformations in terms of the light-like operators
we get the simple form
\begin{align}
  &\delta_\eta b_{\vec p}=-\eta |\vec p| \alpha_{\vec p},
  &&\delta_\eta {b_{\vec p}}^\dagger=\eta |\vec p| {\alpha_{\vec p}}^\dagger,
  \cr
  &\delta_\eta \overline b_{\vec p}=0,
&&\delta_\eta{\overline b_{\vec p}}^\dagger=0,
  \cr
  &\delta_\eta \alpha_{\vec p}=0,
  &&\delta_\eta {\alpha_{\vec p}}^\dagger=0,\cr
  &\delta_\eta \overline \alpha_{\vec p}= \eta |\vec p| \overline b_{\vec p},
  &&\delta_\eta {\overline \alpha_{\vec p}}^\dagger= \eta |\vec p|{\overline b_{\vec p}}^\dagger,
\end{align}
which exhibits the symmetric role played by the operators of the ghosts
$\alpha_{\vec p},\overline\alpha_{\vec p}$
and those of the light-like modes of the electromagnetic field
$b_{\vec p},\overline b_{\vec p}$. Finally, the BRST charge is
$$Q_{\rm BRST}=\int |\vec p|\left(
\overline b_{\vec p} {\alpha_{\vec p}}^\dagger
+
\alpha_{\vec p}{\overline b_{\vec p}}^\dagger
\right)
\frac{\dd^3p}{(2\pi)^3}.
$$

Observe that in terms of these modes the commutation relations and the
BRST transformations are independent of the gauge fixing parameter $\xi$.
One is tempted to say that the latter has disappeared completely from the theory,
which would contradict the fact that the propagator of the gauge field
actually depends on $\xi$.  

The apparent paradox is solved by noticing that in order to compute the propagator
we must know the evolution of the operators and for that we also need the
Hamiltonian. It is in the expression of the latter where the parameter $\xi$ still appears.
In fact, one can verify
 \begin{eqnarray*}
   H_{gh}&=&\int|\vec p|(
   \overline\alpha^\dagger_{\vec p}\alpha_{\vec p})
   +\alpha^\dagger_{\vec p}\overline\alpha_{\vec p}
   )\frac{\dd^3 p}{(2\pi)^3}\cr
   H_{em}&=&\int|\vec p|({a^{(1)}_{\vec p}}^\dagger a^{(1)}_{\vec p}+
 {a^{(2)}_{\vec p}}^\dagger a^{(2)}_{\vec p}
+{\overline b_{\vec p}}^\dagger b_{\vec p}
 +{b_{\vec p}}^\dagger \overline b_{\vec p}
 +\frac{1-\xi}2 {\overline b_{\vec p}}^\dagger \overline b_{\vec p})
 \frac{\dd^3 p}{(2\pi)^3}
 \end{eqnarray*} 
and the last term in the integrand of $H_{em}$
gives rise  to the $\xi$ dependent term of the propagator.
Note in passing that, on the one hand, $[Q_{\rm BRST},H_{em}]+[Q_{\rm BRST},H_{gh}]=0$
and, on the other hand, the $\xi$-dependent term is BRST exact i. e.
$$|\vec p|{\overline b_{\vec p}}^\dagger \overline b_{\vec p}
=\{Q_{\rm BRST},\frac12({\overline\alpha_{\vec p}}^\dagger
\overline b_{\vec p}
 + {\overline b_{\vec p}}^\dagger\overline\alpha_{\vec p})\}$$
 and, therefore, vanishes on $\CH_{\rm phys}$. These two properties show, in the most
 simple way, that the dynamics of the theory is well defined on $\CH_{\rm phys}$
 and it is independent of $\xi$ as one should expect.

 \section{Conclusions}

 We have shown that in order to carry out the canonical quantization of the electromagnetic filed in arbitrary $\xi$-gauge one should go beyond the plane wave like solutions and allow for more general ones. Once the appropriate basis of solutions is chosen it is immediate to find the commutation relations of the creation and annihilation operators and the propagator with the correct Feynman's $\ii\epsilon$ prescription.

The singularity that we find for $\xi=-1$ is a consequence of the choice of basic modes, that actually are not defined for that particular value of the gauge parameter. It is somehow reminiscent of the black hole horizon issue, where the singularity of the metric is associated to a bad choice of coordinates.
Exactly as in the previous case, in the so called {\it light-like} basis the singularity disappears and, actually, the dependence on $\xi$ reduces to the coefficient of a BRST exact term in the Hamiltonian. This proves in a very simple way that physical processes are independent of $\xi$.

Despite the fact that no singularity occurs, it is still interesting to consider the special value $\xi=-1$. In this gauge there is a sort of decoupling of chiralities. Indeed, if we define light cone coordinates $x^{\pm}=x^3\pm x^0$, where direction 3 is along the propagation of the gauge field (assumed a plane wave) and also $A^\pm=A_3\pm A_0$, the equations of motion for this particular value of $\xi$ can be written 
$$\partial_{x^+}^2 A^-=0,\quad \partial_{x^-}^2 A^+=0.$$
Notice that both chiralities are not completely decoupled as the
equations are second order.
Given its special properties this chiral gauge ($\xi=-1$)
might lead to some simplification in the computation of
Feynman diagrams.

Another question is whether we can interpret the $\ii\epsilon$ prescription as a deformation of the action. The letter is necessary if we want to implement correctly the quantization by functional integral in Minkowski space.
This can be done for scalar or Dirac fields and is motivated by the need of a damping factor in the Gaussian integrals. We have shown that the recipe given in \cite{Wein} does not work for $\xi<0$, hence the question is if there are some other prescription that works for every $\xi$ and to what extent it is related to the damping property. The ideas of deforming the metric, as discussed in
\cite{CanRai,Ivas1,Ivas2,Ivas3,Viss1,Viss2}, may shed some light on this question; although, in this case,  we might have to pay the price of breaking covariance in the action.
 
\textbf{Acknowledgments:} Research partially supported by grants E21\_17R, DGIID-DGA and PGC2018-095328-B-100, MINECO (Spain)

\end{document}